# The Laniakea supercluster of galaxies

R. Brent Tully[1], Helene Courtois[2], Yehuda Hoffman[3] & Daniel Pomarède[4]

[1] Institute for Astronomy, University of Hawaii, Honolulu, Hawaii 96822, USA
[2] Universite Claude Bernard Lyon I, Institut de Physique Nucleaire, Lyon, France
[3] Racah Institute of Physics, Hebrew University, Jerusalem 91904, Israel
[4] Institut de Recherche sur les Lois Fondamentales de l'Univers, CEA/Saclay, 91191 Gif-sur-Yvette , France

**Galaxies congregate in clusters and along filaments, and are missing from large regions referred to as voids. These structures are seen in maps derived from spectroscopic surveys[1,2] that reveal networks of structure that are interconnected with no clear boundaries. Extended regions with a high concentration of galaxies are called 'superclusters', although this term is not precise. There is, however, another way to analyse the structure. If the distance to each galaxy from Earth is directly measured, then the peculiar velocity can be derived from the subtraction of the mean cosmic expansion, the product of distance times the Hubble constant, from observed velocity. The peculiar velocity is the line-of-sight departure from the cosmic expansion and arises from gravitational perturbations; a map of peculiar velocities can be translated into a map of the distribution of matter[3]. Here we report a map of structure made using a catalogue of peculiar velocities. We find locations where peculiar velocity flows diverge, as water does at watershed divides, and we trace the surface of divergent points that surrounds us. Within the volume enclosed by this surface, the motions of galaxies are inward after removal of the mean cosmic expansion and long range flows. We define a supercluster to be the volume within such a surface, and so we are defining the extent of our home supercluster, which we call Laniakea.**

**A key component of this paper is an accompanying movie that can be viewed (also in 3D) and downloaded at http://irfu.cea.fr/laniakea or http://vimeo.com/pomarede/laniakea**

The distribution of matter can be alternatively and independently determined from surveys of the distribution of galaxies in projection and redshift or from the motions of galaxies. With the former, using galaxy redshift surveys, the assumption is required that the galaxy lighthouses and mass distribution are strongly correlated, a condition that requires confirmation if, as is suspected, only a minor fraction of matter is baryonic. Moreover with the former there is a stringent demand that the survey be complete, or at least that its incompleteness be well understood. With the latter, studies of galaxy motions, sparse-sampling is acceptable (indeed inevitable) but dealing with errors is a challenge. Except for the very closest galaxies, uncertainties in distance measurements translate into uncertainties in the peculiar velocities of galaxies that are larger in amplitude than the actual peculiar velocities. Many measures are required for suitable averaging and care must be taken to avoid systematics. Overall, the two paths to define the distribution of matter are in good agreement, a circumstance that represents a considerable success for the standard model of structure formation via gravitational instability[4,5,6,7].

The path from velocities to mass distributions benefits from the coherence in velocities on large scales. Multipole components in the velocity field can point to tidal influences beyond the survey region. The current all-sky redshift surveys and distance measurement surveys reach similar depths but the latter probe structure to greater distances because of sensitivity to uncharted attractors and repellers. Coherence in motions on large scales means that signals can be measured by averaging over data in circumstances where individual contributions are very noisy.

Details about the actual measurement of galaxy distances and derivative peculiar velocities are relegated to the Methods section. These two parameters are available for over 8000 galaxies, affording extremely detailed information locally, degrading outward to increasingly coarse coverage. The discussion next will be about a way to use this material to reconstruct the large scale structure of the nearby universe[7].

The underlying 3D velocity and density fields are obtained by the Wiener Filter algorithm[8,9] assuming the standard model of cosmology as a Bayesian prior. Large-scale structure is assumed to develop from gravitational instabilities out of primordial random Gaussian fluctuations. The developing density and velocity fields retain their Gaussian properties as long as the growth is in the linear regime. It has been shown[8] that with a random Gaussian field the optimal Bayesian estimator of the field given the data is the Wiener Filter minimal variance estimator. At the present epoch large-scale structure has become non-linear on small scales. However it is an attractive feature of the velocity field that the break from linearity is only on scales of a few Mpc, an order of magnitude smaller in scale that for the density field. In any event, the present discussion concerns structure on scales of tens to hundreds of Mpc, comfortably in the linear regime.

The Wiener Filter result is determined by the ratio of power to (power + noise). Hence, the large scale structure is strongly constrained nearby where uncertainties are small and the coverage is extensive. At large distances where the data becomes more sparse and noisy the Wiener Filter attenuates the recovered density and velocity fields to the null field that is expected in the absence of data. However in the linear regime there is coherence in galaxy flows on much larger scales than seen in density fluctuations. Tidal influences from beyond the surveyed regions can be manifested in cosmic flows on scales that exceed the coverage in measured distances by a factor two[10].

An ultimate goal is to map the velocity field to a radius that completely encompasses the sources of the motion of the Local Group of 631 km s$^{-1}$ reflected in the cosmic microwave background dipole[11]. However, our knowledge of flows on large scales remains inevitably modulated by the extent of the data. Our analysis of the Cosmicflows-2 data tells us that a coherent flow extends across the full extent of the region that we can map, just reaching the Shapley Concentration[12]. It is clear that we do not yet have a sufficiently extensive compendium of distances to bound the full source of our deviant motion from the cosmic expansion.

For the present discussion the focus is on intermediate scale flow patterns. The standard model of cosmology predicts that on the scale we are considering the flow is irrotational, namely the velocity field is a gradient of a potential, $\vec{v} = -\nabla\phi$ where the velocity potential equals the gravitational potential mediated by a linear factor that depends on cosmological parameters. The local minima and maxima of the potential (attractors and repellers respectively) are the drivers of the large scale flow. We can define a 'basin of attraction' as the volume containing all points whose flow lines converge at a given attractor. The large scale structure can be characterized on scales of a few Mpc and above by attractors and their basins of attractions.

The Wiener Filter provides a straightforward way of decomposing the velocity field into a local component that is induced by the distribution within a zone and a tidal residual[13]. In the linear regime the velocity and density fields, **v** and $\delta$, are directly related: $\nabla\cdot\mathbf{v} = -H_0 f(\Omega_m, \Omega_\Lambda) \delta$ where $f$ depend on the cosmological matter and vacuum energy densities characterized by $\Omega_m$ and $\Omega_\Lambda$. Specify a center and radius and the density field within the defined volume as a cut out from the full Wiener Filter density map. Then the Poisson-like equation between velocity and density can be solved to derive the velocity field responding to just the matter within the prescribed volume. The vector subtraction of the local velocity component from the full flow gives the external component of the velocity field. The residual component is responsible for the bulk motion of the zone under consideration and a quadrupole component within the zone. The decomposition allows us to probe the local velocity field, with the tidal field induced by distant structures filtered out. Relative attractors and their basins of attraction are defined with respect to that local field.

One more useful tool is to be mentioned before turning to results. At each position in space the three eigenvalues of the velocity shear tensor can be calculated. If these eigenvalues are ordered from most positive to most negative then a threshold can be set that captures four possibilities. Flows can be inward on all three axes, the condition of a cluster, inward on two axes and outward on the third, the condition of a filament, inward on one axis and outward on two, hence a sheet, or outward on all three axes, hence a void. Boundaries can be created around contiguous regions with the same shear properties and the contours outline the cosmic web as reconstructed by the V-web algorithm[14].

Only two figures illustrate the discussion on these pages but a fuller presentation is given by the video and additional figures included with the supplemental material. In turn attention is given to smaller (but still linear regime) scales to examine the separation of local and tidal flows and to isolate local basins of attraction. Particular attention is given to locations where there are local divergences. In co-moving coordinates and with the removal of long range flows, there are places where relatively neighboring galaxies can be found to be moving in opposing directions toward separate local basins of attraction. Voids are usually the demarcations between attraction basins but divergences can occur along filaments and sheets. We give particular attention to the very nearby case between our home basin of attraction and the Perseus-Pisces complex[15,16]. There is a particularly impressive example of an apparent bridge between attractors that is diverging given by the feature called the "arch" in a supplemental figure. Similar structures abound with close inspection. Velocity information reveals the divergence locations along filaments between high density regions. This dissipation of the cosmic web is expected with the accelerated expansion of the universe. We are emphasizing that peculiar velocity information can reveal details that are otherwise hard to discern.

The particular interest with the present discussion is with the largest structure that can be circumscribed with the presently available distance and peculiar velocity data. This is the structure schematically illustrated in Figures 1 and 2. The region includes 13 Abell clusters (with the Virgo Cluster). Local flows within the region converge toward the Norma and Centaurus clusters in good approximation to the location of what has been called the 'Great Attractor'[17]. This volume includes the historical Local and Southern superclusters[18], the important Pavo-Indus filament, an extension to the Ophiuchus Cluster, the Local Void, and the Sculptor and other bounding voids. This region of inflow toward a local basin of attraction can be reasonably called a supercluster. The region, if approximated as round, has a diameter 12,000 km s$^{-1}$ or 160 Mpc and encompasses $\sim 1 \times 10^{17} M_\odot$. The region deserves a name. In the Hawaiian language "lani" means "heaven" and "akea" means "spacious, immeasurable". We propose that we live in the Laniakea Supercluster of galaxies.

**References for Main Text**

**Acknowledgements**

We thank our many collaborators in the accumulation of Cosmicflows-2 distances. We thank the CLUES collaboration, and in particular S. Gottloeber and J. Sorce in connection with the analysis. T. Jarrett provided an unpublished 2MASS Extended Source Catalog redshift compendium, the only all-sky redshift catalogue extensive enough to match the region of our reconstruction. The narration in the Supplementary Video is by S. Anvar and the original music is played by N.E. Pomarède. The name 'Laniakea' was suggested by N. Napoleon, Kapiolani Community College, Hawaii. Financial support was provided by US National Science Foundation award AST09-08846, several awards through the Space Telescope Science Institute for observing time with Hubble Space Telescope, an award from the Jet Propulsion Lab for observations with Spitzer Space Telescope, and NASA award NNX12AE70Gfor analysis of data from the Wide-field Infrared Survey Explorer. We also acknowledge support from the Israel Science Foundation (1013/12) and the Lyon Institute of Origins under grant ANR-10-LABX-66 and the CNRS under PICS-06233.

**Author Contributions** R.B.T. guided the project, was involved in the data acquisition, interacted closely in the development of the ideas that are presented here, and wrote most of the Letter. H.C. was involved in the observing programme, was instrumental in coordinating activities, and was involved in all facets. Y.H. took responsibility for the theoretical analysis, including the Wiener filter, the cosmic web and the Malmquist bias correction. D.P. developed and applied visualization tools useful to this research.

**Figure for Main Text**

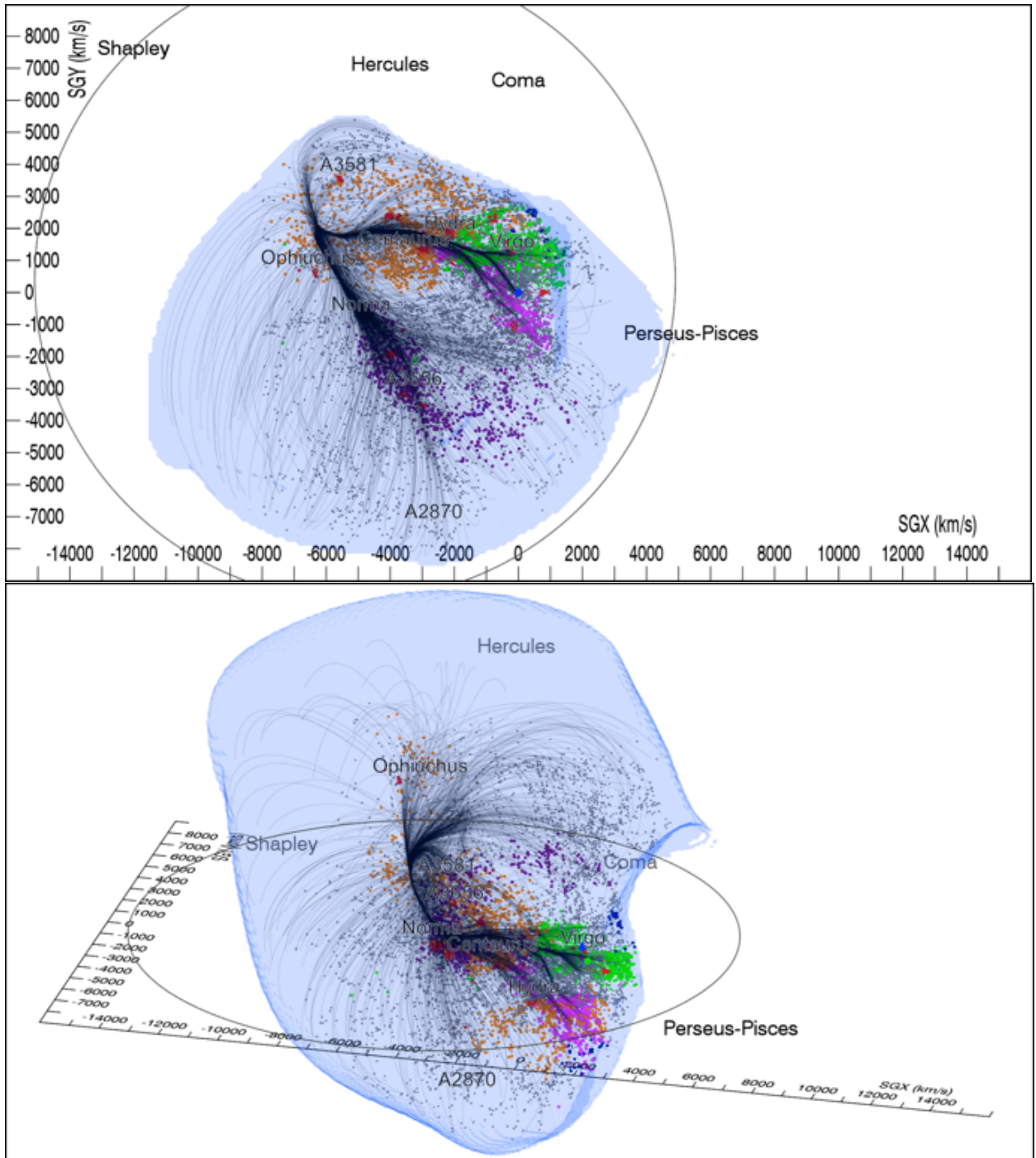

Fig. 1.— Two views of the Laniakea Supercluster. The outer surface demarcates the limits of local velocity flows. Velocity streamlines are shown in black and terminate in the vicinity of the Norma Cluster. Individual galaxies from a redshift catalog are given colors to distinguish major components within the Laniakea Supercluster: the historical Local Supercluster in green, the Great Attractor region in orange, the Pavo-Indus filament in purple, and structures including the Antlia Wall and Fornax-Eridanus cloud in magenta.

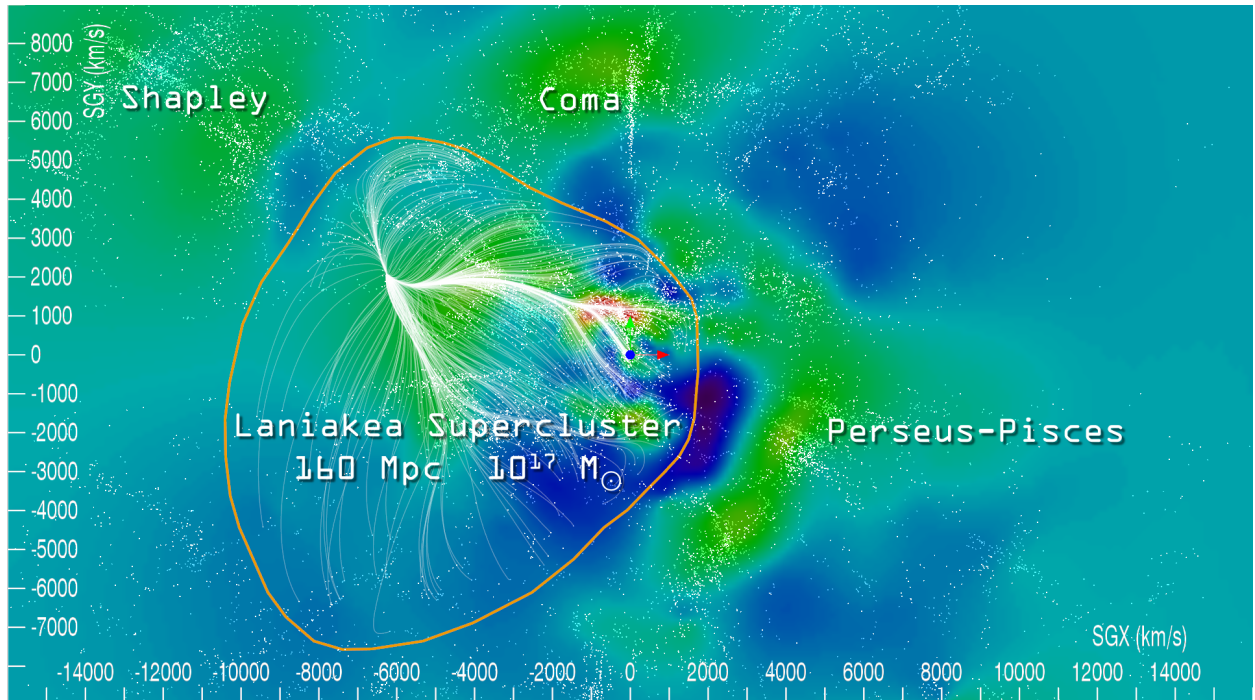

Fig. 2.— A slice of the Laniakea Supercluster in the supergalactic equatorial plane. Shaded contours represent density values within the equatorial slice with red at high densities and blue in voids. Individual galaxies from a redshift catalog are given as white dots. Velocity flow streams within the Laniakea basin of attraction are shown in white. The orange contour encloses the outer limits of these streams. This domain has a extent of ∼12,000 km s$^{-1}$(∼160 Mpc diameter) and encloses ∼$10^{17}$ M$_\odot$.

**Methods Section**

The present discussion draws on a new catalog of galaxy distances and peculiar velocities[1], one that extends to recession velocities of 30,000 km s$^{-1}$ (redshift z = 0.1) and with 8,161 entries provides a density of coverage previously unknown. The new catalog is called Cosmicflows-2 and the six main methodologies for the distance estimates rely on the characteristics of Cepheid pulsations, the luminosity terminus at the tip of the Red Giant Branch, surface brightness fluctuations in elliptical galaxies, the standard candle nature of supernovae of type Ia, the adherence by elliptical galaxies to a fundamental plane in luminosity, radius, and velocity dispersion, and the correlation between the luminosity of spirals and their rate of rotation. Each of the methodologies has strengths and weaknesses. The Cepheid and tip of the Red Giant Branch techniques provide high precision distances but only to very nearby galaxies. The elliptical fundamental plane and spiral luminosity-rotation methods provide individually less accurate distances but can be used to acquire samples of thousands of galaxies in a large volume. Supernovae Ia are excellent distance indicators and can be seen far away but they arise serendipitously and current samples are small. Jointly the sky coverage is now substantial within ∼100 Mpc, with only spotty coverage out to ∼400 Mpc.

A considerable effort with the compilation of the Cosmicflows-2 compendium went into assuring that the six independent methodologies are on a common scale. This scale determines the value of the Hubble Constant. In Cosmicflows-2, this parameter is determined from the velocities and distances of supernovae Ia at redshifts 0.03 < z < 0.5, in a domain where peculiar velocities should be a negligible fraction of observed velocities. We found[2] $H_0 = 75.2 \pm 3.0$ km s$^{-1}$ Mpc$^{-1}$. The value of $H_0$ remains somewhat contentious, especially with the claimed value of $67.3 \pm 1.2$ by the Planck Collaboration[3]. Other determinations that actually measure distances and velocities are more compatible with our value[4,5]. Variations of the assumed $H_0$ within the Cosmicflows-2 analysis introduce a monopole term: infall if $H_0$ is increased and outflow if $H_0$ is decreased. Our value minimizes the monopole term. Tests with the current distances demonstrate unphysical monopole terms if $H_0$ is varied by more than $\pm 1.5$ km s$^{-1}$ Mpc$^{-1}$ from the fiducial value. The Planck determination at face value introduces a massive outflow. It is to be emphasized that, for a study such as ours for galaxy flows, the value assumed for $H_0$ needs to be internally consistent with the scaling of distance measures. A rescaling, say by a re-evaluation of the distances to the nearest galaxies, leaves peculiar velocities unchanged. The extremely low $H_0$ value found by the Planck collaboration is only plausible in the face of the monopole implication if distances to nearby calibrator galaxies are significantly greater than currently measured.

The Wiener Filter was used before to reconstruct the underlying flow field from sparse and noisy velocity surveys[6,7]. Here it is used to overcome the Malmquist bias introduced by errors in distance that scatter galaxies out of regions prominent in galaxies towards emptier regions[8]. This correction is done in two steps. In the first the Wiener Filter is applied to the raw data. New estimated distances are defined by $d_{WF} = (V_{obs} - V_{WF})/H_0$ where $V_{WF}$ is the radial component of the Wiener Filter velocity field at the measured position. This adjustment brings the estimated distances to be much closer to their position in redshift space. The Wiener Filter is then applied to the revised catalog to yield the reconstructed flow that is presented here. The Wiener Filter correction of the Malmquist bias has been tested against mock catalogs drawn from N-body simulations constrained by the Cosmicflows-2 data to reproduce the actual universe. The Wiener Filter strongly suppresses the spurious (negative) monopole term introduced by the Malmquist bias.

The Wiener Filter provides the Bayesian mean field, given the data, its uncertainties and an assumed prior model. The robustness of the Wiener Filter reconstruction is gauged by sampling the distribution of the residual from the mean field, assuming that the underlying field is Gaussian[9]. To meet this end an ensemble of 20 constrained realizations of the underlying flow has been constructed[7]. A local/tidal decomposition has been applied to a sphere of radius 6000 km s$^{-1}$ = 80 Mpc centered on the putative centroid of the attractor region ([−4700, 1300, −500] km s$^{-1}$). Each one of these realizations has been found to harbor a distinct monolithic over-dense structure that dominates that sphere. Extending the radius of the sphere to a radius of 7000 km s$^{-1}$ and above we find the breakdown of the monolithic structure and the emergence of more than one attractor within the sphere. The ensemble of constrained realizations can be used to assess the statistical significance of our claim for the existence of a basin of attraction on a scale of 6000 km s$^{-1}$ = 80 Mpc. All of the 20 realizations, each consistent with the data and the assumed prior ΛCDM model, reproduce such an attractor.

A superiority of the recovery of structure from peculiar velocities with the Wiener Filter over reconstructions from redshift surveys is the sensitivity to structures in regions that cannot be observed, in particular in the heavily obscured equatorial band of the plane of our galaxy. Indeed a large part of our watershed is hidden by obscuration. The connections between the important Pavo-Indus and Centaurus components of Laniakea, the extension up to Ophiuchus Cluster, the filament from the Antlia Wall that ultimately connects to the Perseus-Pisces complex, and the continuation of the Perseus-Pisces filament beyond Perseus are all examples of hidden structures manifested in flow patterns. The two dominant attractors in Shapley and Lepus are both at low galactic latitudes. A further nice example is the identification of the previously unknown Arch.

We follow with comments about the accompanying video, illustrated with scenes from that video. The video and the figures presented in this paper were created using SDvision[10], an interactive visualization software developed within the framework of IDL Object Graphics. In the Extended Data section there is a transcript of the dialog in versions that include a sound track. The transcript is also provided in the closed caption version.

http://irfu.cea.fr/laniakea
(backup http://vimeo.com/pomarede/laniakea)

The video begins with a display of galaxies from the 2MASS Extended Source Catalog[11] with distances given by measured velocities assuming the Hubble law. This display gives a general idea of the structure in the nearby universe although there is increasing incompletion with distance from our position at the center and there is some inhomogeneity in coverage with direction.

During a second half rotation of the scene there is a display of the peculiar velocity vectors that are fundamental to the current analysis. Peculiar motions toward us are in blue and those away are in red. The density of measurements is dense within 7500 km s$^{-1}$ = 100 Mpc and falls off quickly beyond that distance.

The rotations continue but next with the full velocity flows derived from the Wiener Filter analysis. The immediate product of the filtering is a most probable three-dimensional peculiar velocity at each position within the volume of the study. An imaginary vehicle that starts from any seed location is passed to an adjacent position by the locally inferred vector of motion, then again passed onward, creating flows that are captured by the visualized stream lines. The dominant flow is toward the Shapley concentration of galaxies. A second feature involves an accumulation of flows in the Perseus-Pisces region that then streams down to the Lepus region. After a rotation, the implied large scale potential field is superimposed. If the viewing box is extended, it is reasonably convincing that the flow to Lepus continues on to connect with Shapley. In other words, essentially the entire volume being displayed is involved in a flow toward Shapley. This proposition will be explored in a future publication.

The emphasis of the present discussion is on intermediate scales that are within the domain of most of the distance measurements. The contours that become superimposed on the flow lines represent the density field derived from the Wiener Filter velocity field. The amplitudes of the density reconstructions are high near the center of the cube where data constraints dominate errors and taper to low values at the edges where uncertainties are large and the derived density field approaches the mean field. Figure ED-1 in the Extended Data is extracted from this sequence.

After some rotations, the flow lines are seen to break up into disjoint pieces. These disconnections occur because we have begun to use the trick of separating into local and tidal components.

Next there is a transition to the V-web representation. The eigenvalues of the velocity shear can be evaluated at each local position. Surfaces are shown that enclose clusters, filaments, and sheets. Now there is a zoom inward and flow lines are shown for local motions within a zone extending to 6000 km s$^{-1}$ = 80 Mpc from our central position. We are sampling two separate basins of attraction. The red flow lines are directed toward the Perseus-Pisces complex while the black flow lines are directed toward the Centaurus-Norma-Pavo-Indus structures. We lie within the domain of the black flow lines although we are near the boundary where local flow directions flip. It is seen that the flows converge to follow along filaments. There are interesting places along the V-web filaments where local peculiar velocities flip in direction. After some time in this video sequence there is the superposition of individual galaxies from the redshift catalog. Colors are given to members of different structures. The positions of the individual galaxies are not corrected for redshift distortion but reasonable agreement is seen between the locations of galaxies, the flow lines, and the inferred V-web structures. Figure ED-2 is extracted from the video as an illustration of this point in the discussion.

In the concluding section, the local/tidal decomposition is shifted to a center near to the Norma Cluster in order to isolate our basin of attraction. The flow lines now only include those that are moving inward. A surface is created at the boundary of this region. Figure 1 shows two projections from video frames and Figure 2, the concluding frame in the video, is a projection onto the equatorial plane in supergalactic coordinates. This latter figure is augmented in Extended Data Figure ED-3a to include flow lines away, as well as toward, the local region of attraction. In addition there is an indication in the Extended Data figure of the swath of the obscuration in the plane of the Milky Way. The two other panels of Figure ED-3 give two orthogonal views of the inward and outward local flows.

The orange outline is the projected boundary of our local basin of attraction that we call the Laniakea Supercluster. This domain of diameter 160 Mpc, given the mean density of the universe, encloses $10^{17}$ M$_\odot$. We stress that in the fullness of distance measurements on a much larger scale it will almost certainly be demonstrated that Laniakea is not at rest with respect to the cosmic expansion and is only a part of something very large that is at rest in the cosmic reference frame. Those unfamiliar with the field are to be reminded again that cosmic expansion velocities are removed in deriving peculiar motions. Infalling motions on large scales are only perturbations. All galaxies except those in the immediate vicinity of clusters and groups are flying apart.

**References for Methods Section**

**Supplemental Information**

The figures in this article are frame extracts from an accompanying video. The video runs 7 minutes and a soundtrack monolog provides a description. A closed caption option complements the audio (select the "CC" option button).

Transcript of the dialog:

00:08 Let's embark on the discovery of our home supercluster: "Laniakea" All coloured dots in this scene are galaxies belonging to this large structure, as seen from earth. ... We live in a beautiful large blue spiral galaxy ... just one of many in this small universe volume of size 32,000 km/s on a side.
00:35 Each black dot is a galaxy with a measured redshift in the 2MASS Extended Source Catalog. The flux limited selection of this survey results in a spatial coverage, that is densest near the center. As we just saw during the zoom out, obscuration by the Milky Way creates the large vertical artificially vacant zone. Redshift space distortion causes clusters of galaxies to appear elongated, towards the center of the cube.
01:11 Cosmicflows-2 peculiar velocity measurements are superimposed during a second half rotation. Peculiar motions toward us are in blue and motions away are in red.
01:29 The scene now transitions to peculiar velocity flow lines derived from a Wiener Filter analysis of the Cosmicflows-2 peculiar velocity field. .... Our Galaxy is located at the origin indicated by the three central colored directional vectors.
01:54 Here we superimpose the potential field responsible for the peculiar velocity flow. The Shapley galaxy concentration is the dominant local feature. The secondary feature in the direction of Lepus appears to be connected.

02:10 Contours transition to the Wiener Filter over-density field. Contours are prominent near the center where the peculiar velocity information is dense and accurate but diminishes at large distance from the center where information is sparse and uncertain.

02:49 Local components of velocity flows begin to be distinguished. Hercules separates from Shapley and Perseus-Pisces separates from Lepus.

03:16 Here we see contours of the cosmic web as defined by the Velocity-web. Local eigenvalues of the shear tensor are evaluated giving the red, blue, and white contours enclosing knots, filaments, and sheets.

03:35 We see local velocity flows in two adjacent regions. We live in the region of the black flow lines but we are near the transition to the red flow lines associated with the Perseus-Pisces basin of attraction.

04:13 Galaxies from a redshift catalog are superimposed with colors associating them with separate major features.

04:49 The black circle centered at our location defines the region of densities giving rise to these local cosmic flows represented in black and red.

05:03 We now impose a new center and radius on the Wiener Filter density field, and the velocity flow lines are restricted to the watershed of this basin of attraction.

05:13 Galaxies from the redshift catalog are again superimposed with colors distinguishing separate major features. A surface encloses the limits of the watershed.

05:48 Black vectors illustrate the flow of galaxies toward a location near the Norma Cluster, and the local potential minimum of this basin of attraction.

06:24 We call the watershed feeding our basin of attraction the Laniakea Supercluster.

**Extended data Figures**

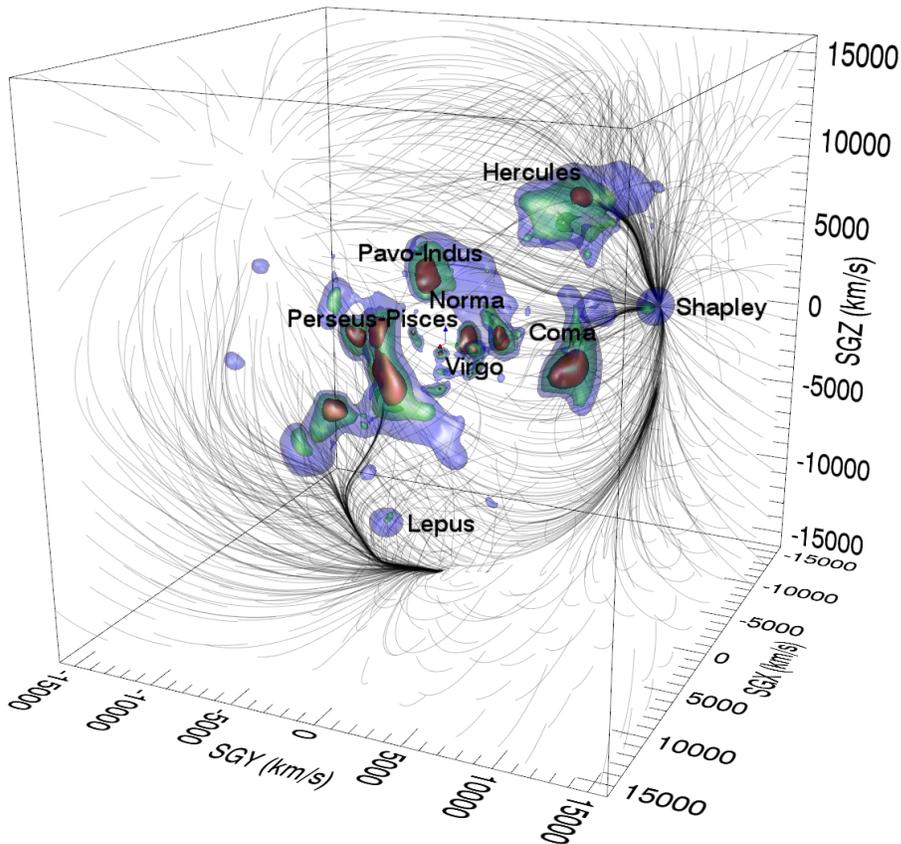

Fig. ED 1. — Structure within a cube extending 16,000 km s$^{-1}$ (~200 Mpc) on the cardinal axes from our position at the origin. Densities on a grid within the volume are determined from a Wiener Filter reconstruction based on the observed velocity field. Three isodensity contours are shown. The density map is detailed near the center of the box where observational constraints are dense and accurate but tapers to the mean density as constraints weaken. Nevertheless, velocity flows illustrated by the black threads are defined on large scales. Ultimately all flows appear to drain toward Shapley although flows through the Perseus-Pisces filament take a circuitous route through the poorly studied Lepus region.

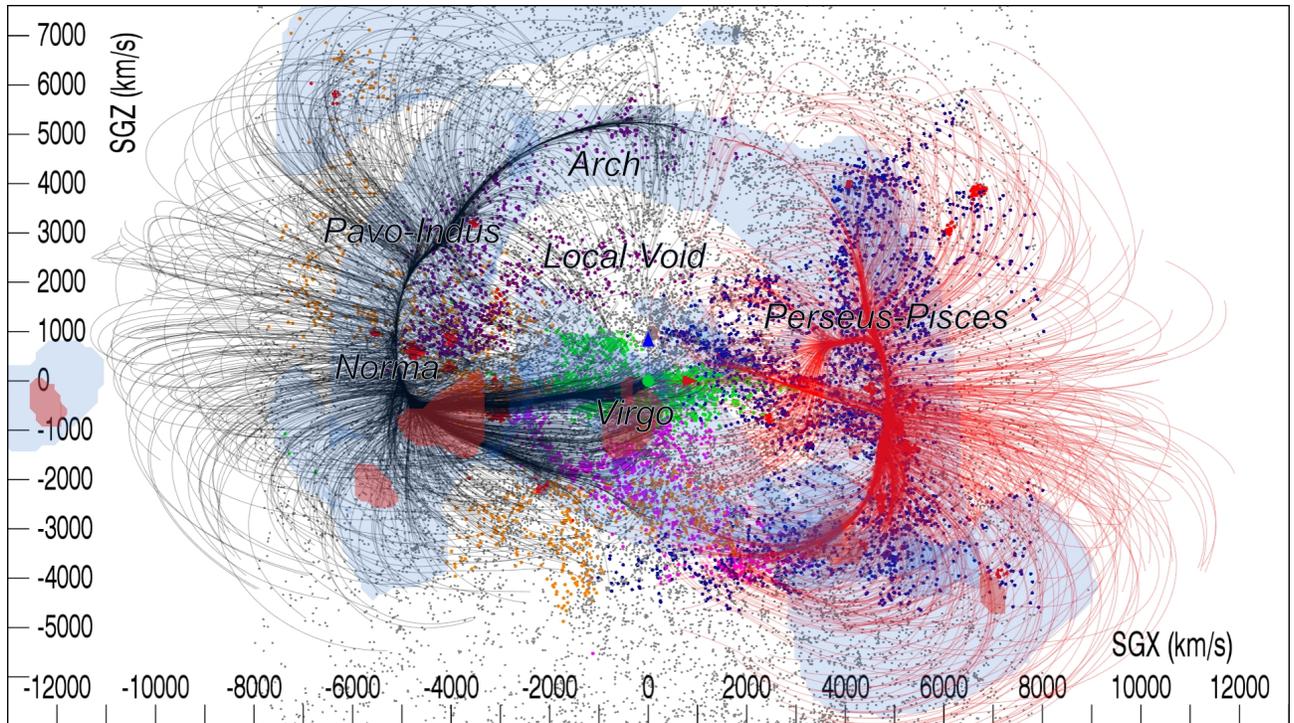

Fig. ED 2. — A representation of structure and flows due to mass within 6,000 km s$^{-1}$ (~80 Mpc). Surfaces of red and blue respectively represent outer contours of clusters and filaments as defined by the local eigenvalues of the velocity shear tensor determined from the Wiener Filter analysis. Flow threads originating in our basin of attraction that terminate near the Norma Cluster are in black and adjacent flow threads that terminate at the relative attractor near the Perseus Cluster are in red. The Arch and extended Antlia Wall structures bridge between the two attraction basins.

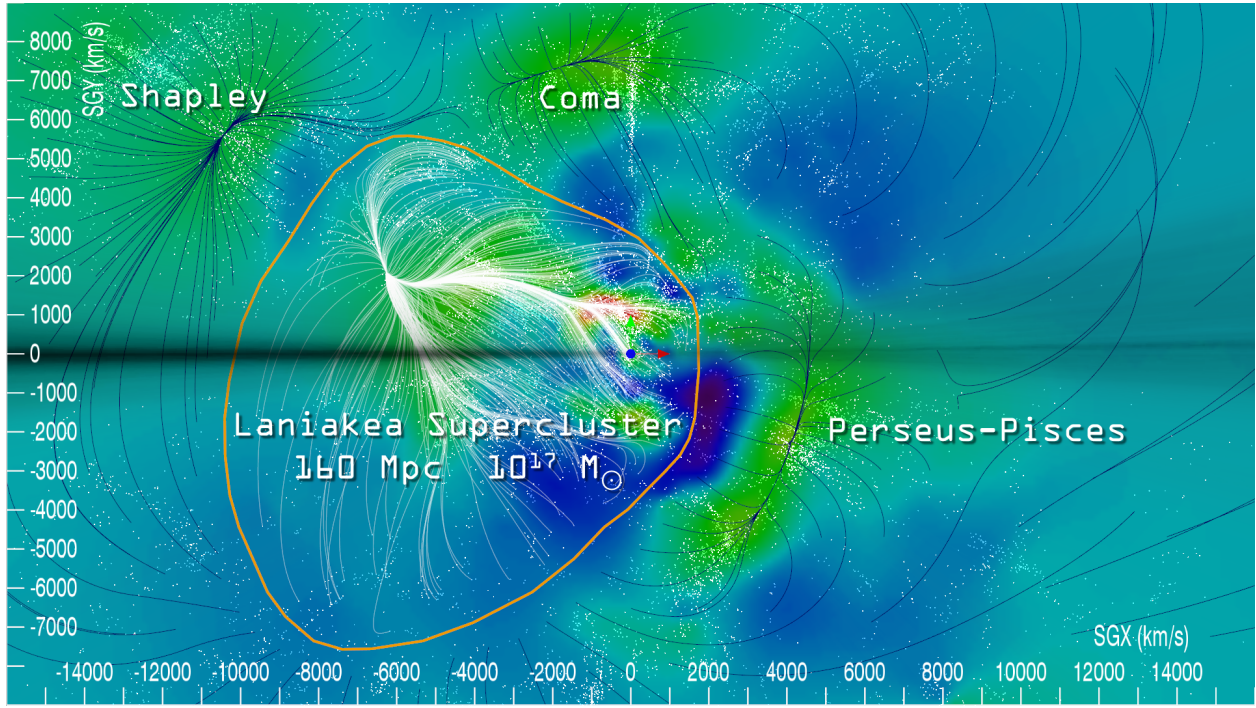
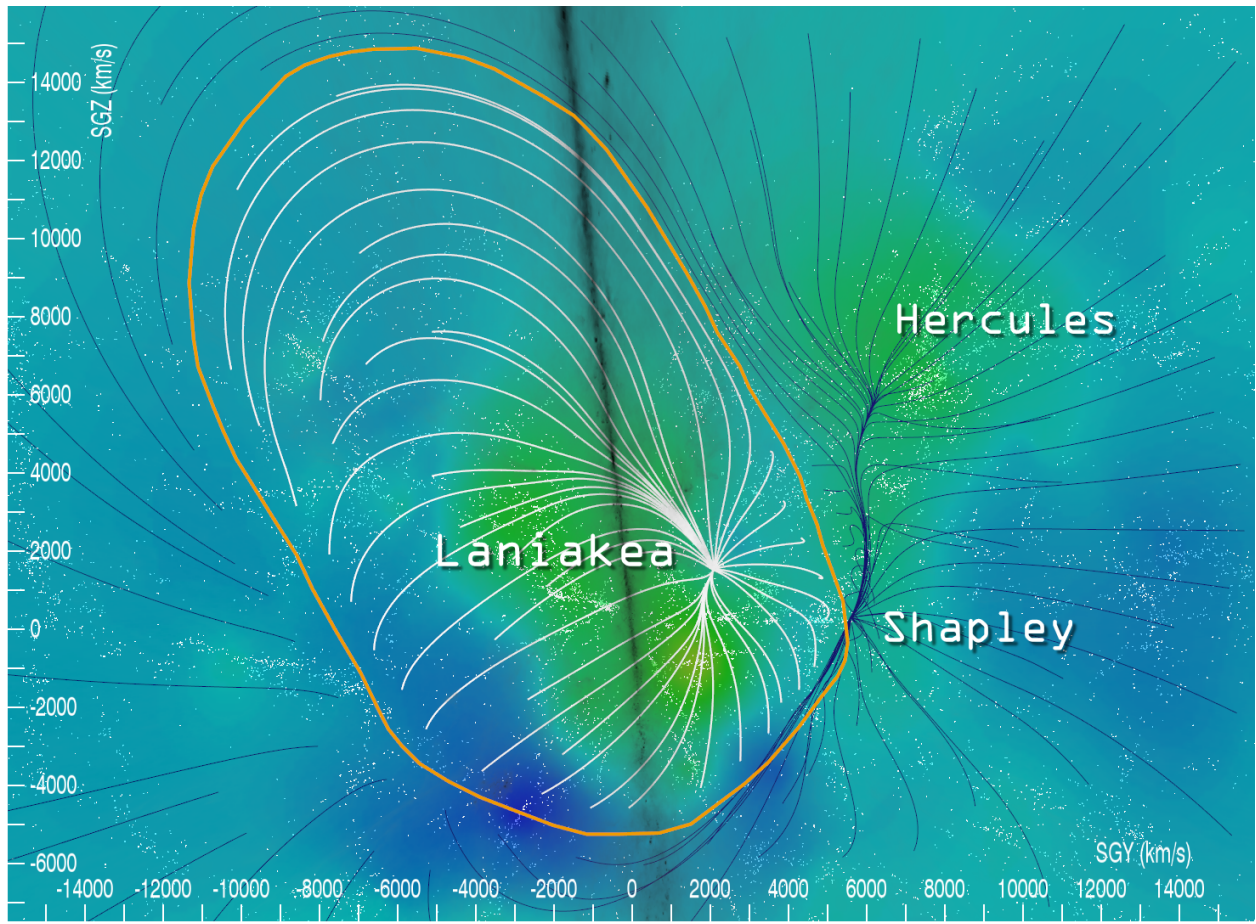

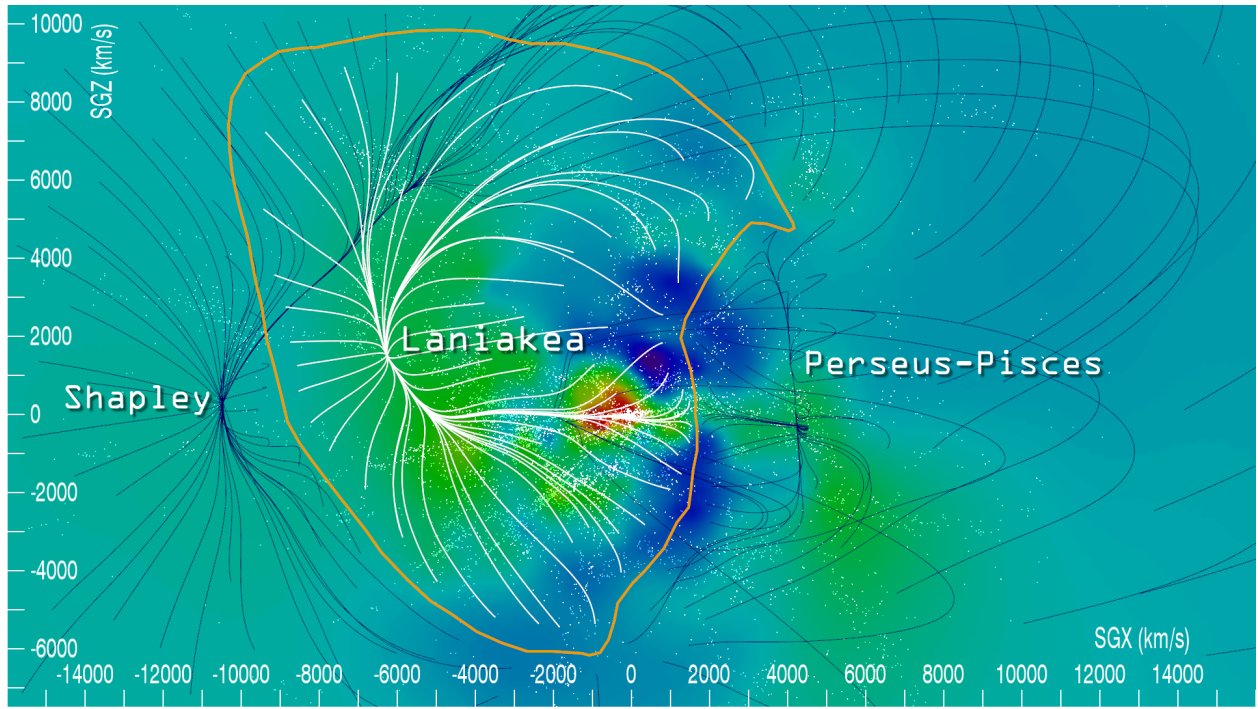

Fig. ED 3. --- Three orthogonal views that illustrate the limits of the Laniakea Supercluster. Panel *a* extends on the scene shown in Fig. 2 of the main text with the addition of dark blue flow lines away from the Laniakea local basin of attraction. This new SGX-SGY slice at SGZ=0 also includes a dark swath at SGY=0 showing the region obscured by the plane of the Milky Way. Panel *b* presents an SGY-SGZ slice at SGX=-4750 km s$^{-1}$ while panel *c* is an SGX-SGZ slice at SGY=+1000 km s$^{-1}$. The swath of Milky Way obscuration is shown in panel *b*. The Milky Way lies essentially in the same plane as the slice in panel *c*, at SGY=0.